\documentclass[twocolumn,english,pratwocolumn]{revtex4}
\usepackage[latin9]{inputenc}
\usepackage{amsmath}
\usepackage{amssymb}
\usepackage{graphicx}

\makeatletter
\@ifundefined{textcolor}{}
{%
 \definecolor{BLACK}{gray}{0}
 \definecolor{WHITE}{gray}{1}
 \definecolor{RED}{rgb}{1,0,0}
 \definecolor{GREEN}{rgb}{0,1,0}
 \definecolor{BLUE}{rgb}{0,0,1}
 \definecolor{CYAN}{cmyk}{1,0,0,0}
 \definecolor{MAGENTA}{cmyk}{0,1,0,0}
 \definecolor{YELLOW}{cmyk}{0,0,1,0}
}

\usepackage{babel}

\makeatother

\usepackage{babel}
\begin{document}

\title{Quantifying Spontaneously Symmetry Breaking of Quantum Many-body
Systems}

\author{G. H. Dong$^{1,2}$, Y. N. Fang$^{1,2,3}$, and C. P. Sun$^{1,2}$}
\email{cpsun@csrc.ac.cn}

\address{$^{1}$Beijing Computational Science Research Center, Beijing 100084,
China\\
 $^{2}$Synergetic Innovation Center of Quantum Information and Quantum
Physics, University of Science and Technology of China, Hefei, Anhui
230026, China\\
 $^{3}$CAS Key Laboratory of Theoretical Physics, Institute of Theoretical
Physics, Chinese Academy of Sciences, and University of the Chinese
Academy of Sciences, Beijing 100190, China}
\begin{abstract}
Spontaneous symmetry breaking is related to the appearance of emergent
phenomena, while a non-vanishing order parameter has been viewed as
the sign of turning into such symmetry breaking phase. Recently, we
have proposed a continuous measure of symmetry of a physical system
using group theoretical approach. Within this framework, we study
the spontaneous symmetry breaking in the conventional superconductor
and Bose-Einstein condensation by showing both the two many body systems
can be mapped into the many spin model. Moreover we also formulate
the underlying relation between the spontaneous symmetry breaking
and the order parameter quantitatively. The degree of symmetry stays
unity in the absence of the two emergent phenomena, while decreases
exponentially at the appearance of the order parameter which indicates
the inextricable relation between the spontaneous symmetry and the
order parameter. 
\end{abstract}
\maketitle

\section{Introduction}

Symmetry and its breaking are evidently of significance in physics.
Many physical laws originate from symmetry. For every continuous symmetry,
it follows from the Noether's theorem \cite{Noether theorem} that
a corresponding conserved law exists; the Kibble-Zurek mechanism \cite{Kibble,Zurek},
on the other hand, allows the dynamical quench through condensed matter
phase transitions to be used as a mean to simulate the formation of
cosmological defects \cite{Kibble-Zurek,Bunkov}. Actually, symmetry
has also been studied in many other subjects such as mathematics \cite{mathmaticas},
biology \cite{biology} and chemistry \cite{chemistry}.

Spontaneous symmetry breaking (SSB) is a fashion of symmetry breaking
of a quantum system S that the Hamiltonian or the motion equation
of S possesses some symmetry while its ground state does not \cite{SSB}.
The importance of SSB is fundamental, as well as practical. For example,
the Higgs mechanism explains the generation of mass for the gauge
bosons in the unified theory for the weak and electromagnetic interactions
\cite{2014 Kibble,1964 Higgs}, and owing to the spontaneous chiral
symmetry breaking in living organism \cite{biology}, synthetic cells
with opposite handedness have been considered as an appealing therapeutic
tool \cite{2016 Wang}. It has been known that the emergent phenomena,
e.g., superconductivity and Bose-Einstein condensation (BEC) \cite{1957 BCS,1956 Onsager}
are all rooted in SSB. The traditional approach to the SSB based emergent
phenomena is the mean field theory (MFT), which is capable of qualitatively
explaining phenomena in diverse areas. A more strict method beyond
MFT was developed by C. N. Yang \cite{ODLRO Yang}, based on the consideration
of off-diagonal long-range order (ODLRO). In ODLRO approach, a non-vanishing
order parameter arises with the emergence of SSB.

Recently, a continuous measure of symmetry breaking has been proposed
by introducing the degree of symmetry (DoS) \cite{2016 Y.N.Fang quantification of nsymmetry}
and then it was applied to the Frobenius-norm-based measures for quantum
coherence and asymmetry \cite{Yao DGH}. Specifically, for a given
transformation set G, the DoS of the Hamiltonian $H$ and the density
matrix $\rho$ of a quantum state are defined, respectively, as follows
\begin{eqnarray}
S(G,H) & = & \frac{1}{4|\tilde{H}|^{2}}\overline{|\{R(g),\tilde{H}\}|^{2}},\label{eq:DoS H}\\
S(G,\rho) & = & \frac{1}{4|\rho|^{2}}\overline{|\{R(g),\rho\}|^{2}},\label{eq:DoS rho}
\end{eqnarray}
where G is a given transformation set, $g\in G$, $R(g)$ is its d-dimensional
representation, $\left|A\right|=\sqrt{\mathrm{Tr}A^{\dagger}A}$,
and $\overline{B(g)}$ is an average defined on G. Especially, $\widetilde{H}=H-\mathrm{Tr}\{H\}\mathrm{I}_{d\times d}/d$
is a re-biased Hamiltonian, which possesses a similar energy spectrum
as that of $H$ but is free of the choice of the energy zero point.
It is worth mentioning that $|\rho|^{2}=\mathrm{Tr}(\rho^{2})$ denotes
the purity of $\rho$. The definition of the DoS satisfies three nice
properties which are physically reasonable: (i) Independent of the
basis in the system's Hilbert space; (ii) Independent of the choice
of the ground state energy; (iii) Scaling invariance \cite{2016 Y.N.Fang quantification of nsymmetry}.

It is well known that the order parameter usually characterises SSB,
and thus it does make sense physically that the order parameter possesses
an underlying connection with the DoS. Actually, one can expect that
the increasing of the order parameter corresponds to the decreasing
of the DoS. The present paper is aimed at establishing a quantitative
relation between the DoS and the order parameter. We consider two
representative phenomena that have been well explored following the
traditional approach, namely the superconductivity with isotropic
pairing and the Bose-Einstein condensation (BEC) in the many spin
model. It will be shown that the DoS for those two types of SSB phenomena
could be expressed in terms of the corresponding order parameters
in the thermodynamic limit. This result represents an important step
to justify the potential of applying this DoS based approach to detect
unknown symmetry breaking related effects in systems whose order parameters
are not awared of in advance. 

In order to calculate the DoS for the conventional superconductor
and BEC, we consider the many spin model with Hamiltonian 
\begin{equation}
H=\sum_{i=1}^{N}\epsilon\sigma_{z}^{i}+\lambda\sigma_{x}^{i}+\mu\sigma_{y}^{i}\label{eq:many spins Hamiltonian}
\end{equation}
where $\epsilon,\lambda$ and $\mu$ are real and $\sigma_{n}^{i}(n=x,y,z)$
denotes the Pauli operator of the $i$th particle. We will show that
both the Bardeen-Cooper-Schrieffer (BCS) and BEC Hamiltonians can
be mapped into Eq.(\ref{eq:many spins Hamiltonian}). Then, the DoS
of the BCS and BEC systems is given through that of the many spin
model. 
\begin{eqnarray}
S(G,\rho_{T=0})_{S} & = & \frac{1}{2}+\frac{1}{2}\exp\left(-(2-\sqrt{2})\pi g(0)\left|\Delta(0)\right|\right)\text{,}\nonumber \\
S(G,\rho_{T=0})_{B} & = & \frac{1}{2}+\frac{1}{2}\exp\left(-2\left\langle a\right\rangle _{0}^{2}\right),\label{eq:conclusion in introduction}
\end{eqnarray}
where the subindices $S$ stands for superconductivity and $B$ for
BEC, $g(\epsilon)$ is the density of states, and $\Delta(0)$ together
with $\left\langle a\right\rangle _{0}$ are the corresponding order
parameters at the absolute zero temperature. It follows from Eq.(\ref{eq:conclusion in introduction})
that the DoS reaches 1 (the symmetry is unbroken) and the symmetry
is totally recovered as $\Delta(0)$ or $\left\langle a\right\rangle _{0}$
vanishes.

This paper is organized as follows: in Sec. II, the mapping from the
BCS and BEC systems to the many spin model is illustrated. In Sec.
III, the DoS for the many spin model is explicitly evaluated and its
relation with SSB is discussed. Then the DoS for the BCS and the BEC
systems are separately explored in Sec. IV and Sec. V, respectively.
Finally, we draw the conclusion in Sec. VI .

\section{Many spin model for the conventional superconductor and BEC}

In the conventional superconductivity theory \cite{1957 BCS}, the
BCS Hamiltonian is obtained by eliminating the phonon variable. 
\begin{equation}
H_{S}=\sum_{k}\epsilon_{k}(a_{k}^{\dagger}a_{k}+b_{k}^{\dagger}b_{k})-V\sum_{k,k^{'}}a_{k}^{\dagger}b_{k}^{\dagger}b_{k^{'}}a_{k^{'}},\label{eq:superconductivity H}
\end{equation}
where $a_{k}(b_{k})$ denotes the annihilation operator of an electron
with momentum $k(-k)$ and spin up (down). In accordance with the
BCS assumption \cite{1957 BCS,1956 Cooper pairs}, the net electron-phonon
attractive interaction $V$ is non-zero only for single electron states
whose energy satisfy $|\varepsilon_{k}-\varepsilon_{F}|\leq\hbar\omega_{D}$,
with $\omega_{D}$ the Debye frequency and $\varepsilon_{F}|$ the
chemical potential in the normal phase. Thus the summation over $k$
in Eq.(\ref{eq:superconductivity H}) is correspondingly restricted
to a thin shell around the sphere whose radius is given by the Fermi
wavevector $k_{F}$. Two electrons with opposite momentum and spin
create an electron pair which is called the Cooper pair \cite{1956 Cooper pairs}.
The Jordan-Wigner transformation \cite{Quantum phase transition}
exactly maps the fermions model into the pseudospins model as 
\begin{eqnarray}
\sigma_{+}^{k}=b_{k}a_{k},\sigma_{-}^{k}=a_{k}^{\dagger}b_{k}^{\dagger},\nonumber \\
\sigma_{z}^{k}=1-(a_{k}^{\dagger}a_{k}+b_{k}^{\dagger}b_{k}).\label{eq:Jordan-Wigner}
\end{eqnarray}

It is easily checked that these pseudospin operators defined above
satisfy the commutation relations of spin type 
\begin{eqnarray}
[\sigma_{+}^{k},\sigma_{-}^{k^{'}}] & = & \sigma_{z}^{k}\delta_{k,k^{'}},\nonumber \\{}
[\sigma_{z}^{k},\sigma_{\pm}^{k^{'}}] & = & \pm2\sigma_{\pm}^{k}\delta_{k,k^{'}}.
\end{eqnarray}
Then, the BCS Hamiltonian given in Eq.(\ref{eq:superconductivity H})
is re-expressed as 
\begin{equation}
H_{S}=-\left(\sum_{k}\epsilon_{k}\sigma_{z}^{k}+V\sum_{k,k^{'}}\sigma_{-}^{k}\sigma_{+}^{k^{'}}\right)+\sum_{k}\epsilon_{k},\label{eq:Superconductivity H with constant}
\end{equation}
where for a given system $\sum_{k}\epsilon_{k}$ is determinate and
can be dropped. We then obtain the BCS Hamiltonian described in the
spin model 
\begin{equation}
H_{S}=-\left(\sum_{k}\epsilon_{k}\sigma_{z}^{k}+V\sum_{k,k^{'}}\sigma_{-}^{k}\sigma_{+}^{k^{'}}\right).
\end{equation}

In the BCS theory, one deal with the eigenenergy problem with the
mean field approximation (MFA), which assumes that the difference
between $\sigma_{-}^{k}(\sigma_{+}^{k^{'}})$ and its expect value
is a small quantity. That is to say, $\sigma_{-}^{k}=\left\langle \sigma_{-}^{k}\right\rangle +\lambda$
and $\sigma_{+}^{k}=\left\langle \sigma_{+}^{k}\right\rangle +\tau$
where $\lambda$ and $\tau$ are small quantities. Then, we make another
assumption that the summation of the averages of $\sigma_{+}^{k}$
is non-zero. 
\begin{eqnarray}
\Delta & = & V\sum_{k}\left\langle \sigma_{+}^{k}\right\rangle ,\label{eq:order parameter}
\end{eqnarray}
where $\Delta$ is the energy gap of BCS. $\Delta$ serves as the
order parameter for the superconducting transition. $\Delta$ is zero
when $T$ is above the critical temperature $T_{c}$, indicating that
the effective interaction is no more attractive. As a result, the
Cooper pairs around the Fermi surface are seperated. When the temperature
is below $T_{c}$, $\Delta$ becomes non-zero, which implies that
the number of electrons is not conserved in $H_{S}$ and the gauge
symmetry $a_{k}(b_{k})\rightarrow a_{k}(b_{k})\exp[i\varphi/2]$ is
spontaneously broken. Then, the BCS Hamiltonian reads

\begin{eqnarray}
H_{S} & = & -\sum_{k}(\epsilon_{k}\sigma_{z}^{k}+\mathrm{Re}(\Delta)\sigma_{x}^{k}+\mathrm{Im}(\Delta)\sigma_{y}^{k}).\label{eq:superconductor H simplified}
\end{eqnarray}
Obviously, the BCS Hamiltonian $H_{S}$ possesses the same form as
the general many spin model Hamiltonian which we present in Eq.(\ref{eq:many spins Hamiltonian}).

For the bosons system, the condensation happens if a finite fraction
of the particles occupies the lowest single-particle state under the
thermodynamic limit \cite{1956 Onsager}. Mathematically, this is
expressed by the appearance of the ODLRO \cite{ODLRO Yang}, i.e.,
\begin{equation}
\rho(x,y)=\langle\hat{\psi}^{\dagger}(x)\hat{\psi}(y)\rangle\overset{\underrightarrow{|x-y|\rightarrow\infty}}{}\langle\hat{\psi}^{\dagger}(x)\rangle\langle\hat{\psi}(y)\rangle\ne0,\label{Onsager criterion}
\end{equation}
where $\rho(x,y)$ is the single particle reduced density matrix,
$\hat{\psi}(x)$ is the bosonic field operator, and the average is
performed under the ground state of the many-body system. $\left\langle \hat{\psi}(x)\right\rangle $
is the order parameter of BEC and BEC occurs when $\left\langle \hat{\psi}(x)\right\rangle $
is non-zero. Actually, the apperance of BEC breaks the $\mathrm{U}(1)$
symmetry which corresponds to the conservation of particle number.
Therefore, the Hamiltonian for BEC is over-simplified as 
\begin{equation}
H\sim a^{\dagger}a+\alpha\left(a+a^{\dagger}\right).\label{eq:BEC coherent}
\end{equation}
The representation of a group element of $\mathrm{U}(1)$ is $R(\theta)=\exp(i\theta a^{\dagger}a)$.
It could be proved that $[R(\theta),H]=0$ if and only if $\alpha=0$.
Actually, the ground state of $H$ is a coherent state $\left|\alpha\right\rangle $
when $\alpha$ is nonzero. According to the Penrose-Onsager criterion
\cite{1956 Onsager}, $\left\langle \alpha\right|a\left|\alpha\right\rangle =\alpha$
is the non-vanishing order parameter.

In order to analyze the DoS of BEC with a general many spin model,
we introduce the many spin model with the Hamiltonian $H_{B}$ 
\begin{equation}
H_{B}=\sum_{i=1}^{N}\epsilon\sigma_{z}^{i}+\lambda\sigma_{x}^{i}.\label{eq:BEC pesudospin}
\end{equation}
By using

\begin{align*}
J_{k} & =\sum_{i=1}^{N}\frac{1}{2}\sigma_{k}^{i}\left(k=x,y,z\right),\\
J_{\pm} & =J_{x}\pm iJ_{y}
\end{align*}
as the collective angular momentum operators, the above Hamiltonian
becomes 
\begin{equation}
H_{B}=2\epsilon J_{z}+\lambda\left(J_{+}+J_{-}\right).\label{eq:H=00003D00003DJ}
\end{equation}
As $\left[J^{2},H_{B}\right]=0$, the total angular momentum conserves.
In the limit of the low excitation, we obtain, 
\begin{align*}
J^{2} & =\frac{N}{2}\left(\frac{N}{2}+1\right),\\
 & =J_{z}^{2}+J_{+}J_{-}-J_{z},
\end{align*}
and 
\[
J_{z}=\frac{1}{2}\pm\frac{N}{2}\sqrt{\left(1+\eta\right)^{2}-4J_{+}J_{-}},
\]
where $\eta=1/N$. Now we map the angular momentum operators to boson
operators in the large $N$ limit, i.e., $a=J_{-}/\sqrt{N},a^{\dagger}=J_{+}/\sqrt{N}$.
Actually, the commutation relation between $a$ and $a^{\dagger}$
is 
\begin{eqnarray}
[a,a^{\dagger}] & = & -\eta\left(1-\frac{1}{\eta}\sqrt{\left(1+\eta\right)^{2}-4\eta a^{\dagger}a}\right),\label{eq:=00003D00005Ba,a+=00003D00005D}
\end{eqnarray}
where we have taken 
\[
J_{z}=\frac{1}{2}-\frac{N}{2}\sqrt{\left(1+\eta\right)^{2}-4\eta a^{\dagger}a}.
\]
When $N\rightarrow\infty$ and $\eta\rightarrow0$, by taking Eq.(\ref{eq:=00003D00005Ba,a+=00003D00005D})
to the first order, we obtain $[a,a^{\dagger}]\simeq1-2\eta a^{\dagger}a\simeq1$.
It is clear that $a(a^{\dagger})$ can be treated as the annihilation
$($creation$)$ operator of bosons in the limit of low excitation
and large $N$. The Hamiltonian $H_{B}$ can be expressed as 
\begin{equation}
H_{B}\simeq2\epsilon a^{\dagger}a+\lambda\sqrt{N}\left(a+a^{\dagger}\right),\label{eq:H=00003D00003Daa+}
\end{equation}
which is just the simplified BEC Hamiltonian we consider in Eq.(\ref{eq:BEC coherent}).
Since we have mapped the many spin model to the bosons model, we make
use of Eq.(\ref{eq:BEC pesudospin}) to simulate SSB in BEC in the
limit of low excitation and large $N$.

In conclusion, we have just showed above that the many spin model
is valid both in fermions and bosons systems.

\section{the dos of the many spin system}

According to the definition of the DoS given in Eqs.(\ref{eq:DoS H},
\ref{eq:DoS rho}), we calculate the DoS of the many spin system as
follows. The density matrix reads $\rho=\exp(-\beta H)/Z,$ where
$Z$ is the partition function $\mathrm{Tr}(\exp(-\beta H))$. Specially,
near the absolute zero temperature, i.e., $\beta\rightarrow\infty$,
the system stays in its ground state. Actually, the linear superposition
of $\sigma_{x},\sigma_{y}$ and $\sigma_{z}$ appeared in Eq.(\ref{eq:many spins Hamiltonian})
can be treated as $\sigma_{z}$ rotated about some certain axis. Thus,
the Hamiltonian appeared in Eq.(\ref{eq:many spins Hamiltonian})
can be simplified to 
\begin{equation}
H=\sum_{i=1}^{N}\xi R_{i}(\theta,\phi)\sigma_{z}^{i}R_{i}^{\dagger}(\theta,\phi),\label{eq:many spins H simplified}
\end{equation}
where 
\begin{align*}
\xi & =\sqrt{\epsilon^{2}+\lambda^{2}+\mu^{2}},\cos\theta=\frac{\epsilon}{\xi},\tan\phi=\frac{\mu}{\lambda},\\
 & R_{i}(\theta,\phi)=e^{-i\phi\sigma_{z}^{i}/2}e^{-i\theta\sigma_{y}^{i}/2}.
\end{align*}
Then, the ground state of this Hamiltonian is obtained immediately
$\left|G\right\rangle =\prod_{i}R_{i}(\theta,\phi)\left|\downarrow\right\rangle _{i},$
where $\left|\downarrow\right\rangle $ denotes the state of spin
down. Here, we regard $\lambda$ and $\mu$ as the pertubations and
$\sigma_{z}^{i}$ remains unchanged under rotations about the $z-$axis
by an arbitrary angle. All of these symmetric transformations form
a group called $\mathrm{SO}(2)$.

In the many spin system here, the symmetric group is $\mathrm{SO}(2)^{\otimes N}$,
with the elements $R(g)=\prod_{i}\exp(-i\omega_{i}\sigma_{z}^{i}/2)$.
The DoS of Hamiltonion is given by 
\begin{eqnarray}
S(\mathrm{SO}(2)^{\otimes N},H) & = & \frac{1}{2}+\frac{\epsilon^{2}}{2\xi^{2}},\label{eq:many spins DoS H}
\end{eqnarray}
where the group average here is $1/(2\pi)^{N}\int_{-\pi}^{\pi}d\omega_{i}...\int_{-\pi}^{\pi}d\omega_{N}$
which means that $N$ particles are rotated separately.

On the other hand, the DoS of the ground state is obtained as 
\begin{eqnarray}
S(\mathrm{SO}(2)^{\otimes N},\rho_{T=0}) & = & \frac{1}{2}+\frac{1}{2}(1-\frac{\lambda^{2}+\mu^{2}}{2\xi^{2}})^{N}.\label{eq:many spins DoS ground state}
\end{eqnarray}
This gives the DoS of the thermal equilibrium state as

\begin{eqnarray}
S(\mathrm{SO}(2)^{\otimes N},\rho_{T}) & = & \frac{1}{2}+\frac{1}{2}(1-\Lambda)^{N},\label{eq:many spins DoS rho}
\end{eqnarray}
where $\Lambda=\left[(\lambda^{2}+\mu^{2})\sinh^{2}(\beta\xi)\right]/\left[\xi^{2}\cosh(2\beta\xi)\right].$

All the Eqs.(\ref{eq:many spins DoS H}, \ref{eq:many spins DoS ground state},
\ref{eq:many spins DoS rho}) decrease as the pertubations $\lambda$
and $\mu$ grow. It means that the symmetry is broken by the pertubations.
Especially, 
\begin{align}
\lim_{\lambda,\mu\rightarrow0}\lim_{N\rightarrow\infty}S(\mathrm{SO}(2)^{\otimes N},H) & =1,\nonumber \\
\lim_{N\rightarrow\infty}\lim_{\lambda,\mu\rightarrow0}S(\mathrm{SO}(2)^{\otimes N},H) & =1.
\end{align}
while at sufficiently low temperature, 
\begin{align}
\lim_{N\rightarrow\infty}\lim_{\lambda,\mu\rightarrow0}S(\mathrm{SO}(2)^{\otimes N},\rho_{T}) & =1,\nonumber \\
\lim_{\lambda,\mu\rightarrow0}\lim_{N\rightarrow\infty}S(\mathrm{SO}(2)^{\otimes N},\rho_{T}) & =\frac{1}{2}.\label{eq:non-commutativity}
\end{align}
The non-commutativity of the limits $N\rightarrow\infty$ and $\lambda,\mu\rightarrow0$
in Eq.(\ref{eq:non-commutativity}) indicates the emergence of SSB
\cite{Kerson Huang}.

\section{Quantifying SSB in fermions system}

The Hamiltonian of the conventional superconductor in BCS theory is
re-expressed as that of the many spin model 
\begin{equation}
H_{S}=-\sum_{k}\xi_{k}R(\theta_{k},\phi)\sigma_{z}^{k}R^{\dagger}(\theta_{k},\phi),
\end{equation}
where 
\begin{align*}
\xi_{k} & =\sqrt{\epsilon_{k}^{2}+\left|\Delta\right|^{2}},\tan\phi=\frac{\mathrm{Im}(\Delta)}{\mathrm{Re}(\Delta)},\tan\theta_{k}=\frac{\left|\Delta\right|}{\epsilon_{k}},\\
 & R(\theta_{k},\phi)=e^{-i\phi\sigma_{z}^{k}/2}e^{-i\theta_{k}\sigma_{y}^{k}/2}.
\end{align*}
The ground state of this superconductor is $\left|G\right\rangle _{S}=\prod_{k}R(\theta_{k},\phi)\left|\uparrow\right\rangle _{k}.$
The quasiparticle excitation energy $\xi_{k}=\sqrt{\epsilon_{k}^{2}+\left|\Delta\right|^{2}}\geq\left|\Delta\right|$.
To excite a quasiparticle around the Fermi surface, one needs at least
an energy scale of $\left|\Delta\right|$, which ensures the stability
of the superconductor. As a consequence, $\left|\Delta\right|$ describes
the energy gap in BCS.

Straightforward calculation shows that the DoS of the Hamiltonian
and the ground state of the conventional superconductor are given
as 
\begin{eqnarray}
S(\mathrm{SO}(2)^{\otimes N},H_{S}) & = & \frac{1}{2}+\frac{1}{2}\frac{\sum_{k}\epsilon_{k}^{2}}{\sum_{k}\xi_{k}^{2}},\label{eq:DoS H fermions}\\
S(\mathrm{SO}(2)^{\otimes N},\left|G\right\rangle _{S}) & = & \frac{1}{2}+\frac{1}{2}\prod_{k}\left(1-\frac{1}{2}\frac{\left|\Delta(0)\right|^{2}}{\epsilon_{k}^{2}+\left|\Delta(0)\right|^{2}}\right)\nonumber \\
 & \simeq & \frac{1}{2}+\frac{1}{2}e^{-\kappa\left|\Delta(0)\right|},\label{eq:DoS GS fermions 2}
\end{eqnarray}
where $\kappa=(2-\sqrt{2})\pi g(0)$ and $g(\epsilon)$ is the density
of states. For more details, see Appendix A. Like Eq.(\ref{eq:many spins DoS ground state})
in the many spin model, the DoS of the ground state in fermions system
also possesses the non-commutativity of two limits. This is one of
the main results of this paper which reflects a direct correspondence
between the DoS and the order parameter. The DoS of the conventional
superconductor is less than unity as long as there exists a non-vanishing
energy gap $\Delta(0)$. It agrees with the fact that SSB occurs when
$\Delta(0)$ is non-zero. The energy gap at the absolute zero temperature
is defined as

\begin{eqnarray}
\Delta(0) & = & V\sum_{k}\left\langle G\right|\sigma_{+}^{k}\left|G\right\rangle _{S}=\frac{V}{2}\sum_{k}\frac{\Delta(0)}{\sqrt{\epsilon_{k}^{2}+\left|\Delta(0)\right|^{2}}}.\nonumber \\
\end{eqnarray}

At a finite temperature T, the density matrix of the system reads
$\rho_{S}^{T}=\exp(-\beta H_{S})/Z$, with the partition function
$Z=4^{N}\prod_{k}\cosh^{2}(\beta\xi_{k}/2)$. The corresponding DoS
of $\rho_{S}^{T}$ is

\begin{eqnarray}
S(\mathrm{SO}(2)^{\otimes N},\rho_{S}^{T}) & = & \frac{1}{2}+\frac{1}{2}\prod_{k}G(\epsilon,\Delta(T)),
\end{eqnarray}
where $G(\epsilon,\Delta(T))=1-\left|\Delta(T)\right|^{2}\tanh^{2}(\beta\xi_{k})/2\xi_{k}^{2}$.
Further simplification shows 
\begin{eqnarray}
S(\mathrm{SO}(2)^{\otimes N},\rho_{S}^{T}) & \simeq & \frac{1}{2}+\frac{1}{2}e{}^{2g(0)\left|\Delta(T)\right|K(T)},\label{eq:DoS BCS rho T}
\end{eqnarray}
where

\begin{align*}
K(T) & =\int_{0}^{\infty}\ln\left[1+\frac{1}{1+t^{2}}(-\frac{1}{2}+\frac{1}{k(T,t)+1})\right]dt,\\
k(T,t) & =\cosh(2\beta\left|\Delta(T)\right|\sqrt{1+t^{2}}),
\end{align*}
and in the above calculation we have assumed $\hbar\omega_{D}\gg\left|\Delta(T)\right|$
for simplicity. It follows from Eq.(\ref{eq:DoS BCS rho T}) that
$K(T)=\ln(2S-1)/\text{(}2g(0)\left|\Delta(T)\right|)$. Hence in fact,
$K(T)\propto\ln(2S-1)/\left|\Delta(T)\right|$.

\begin{figure}
\centering{}\centering{}\includegraphics[scale=0.33]{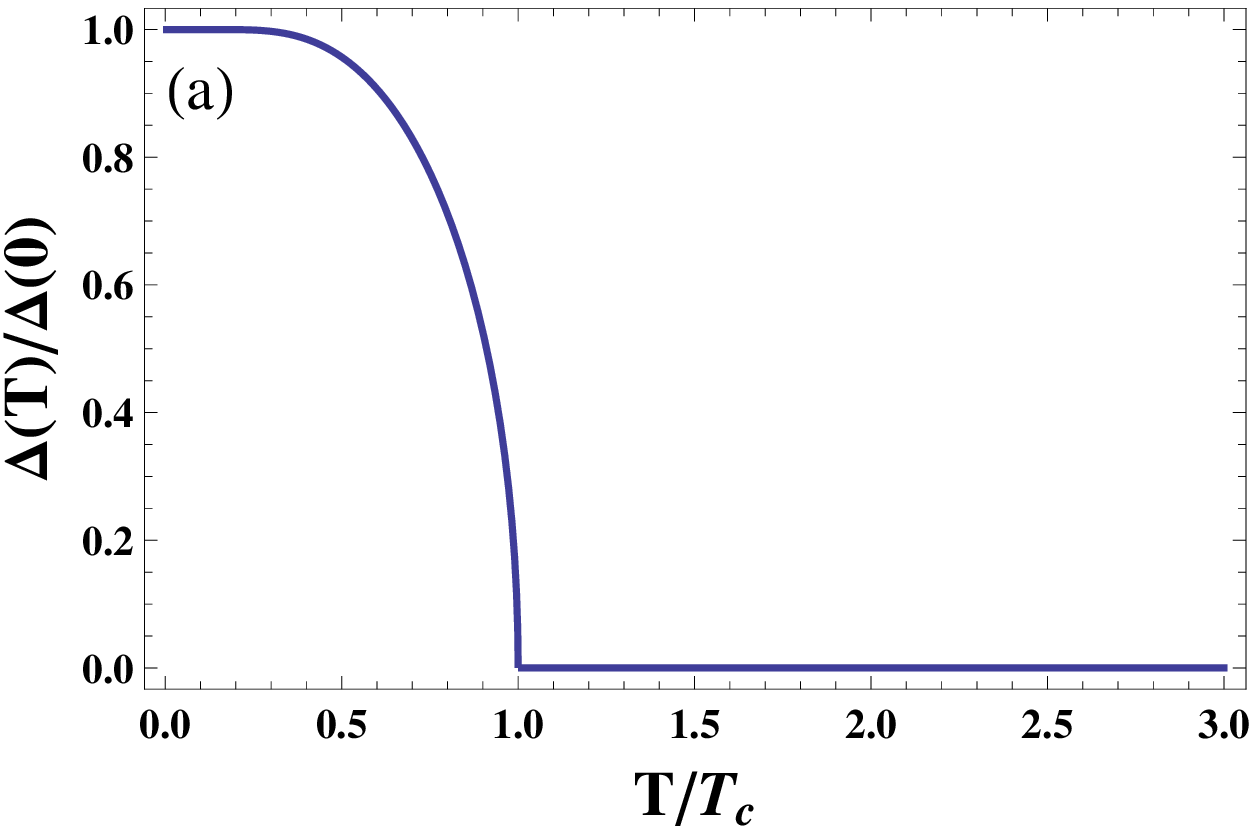}\includegraphics[scale=0.45]{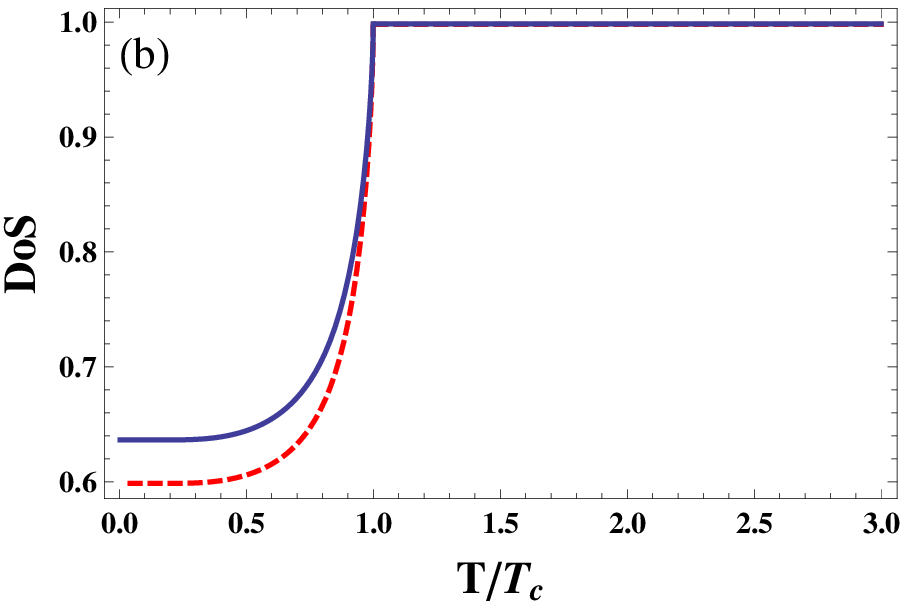}\caption{(a) Energy gap $\Delta$ vs temperature $T$ for the conventional
superconductor. The energy gap is normalized to its value at the zero
temperature while $T$ is measured in the unit of the critical temperature
$T_{c}$. (b) The degree of symmetry of $\rho_{S}^{T}$ with respect
to the $\mathrm{\mathrm{SO}}(2)$ transformation group. The blue solid
and red dashed curves are corresponded to $g(0)k_{B}T_{c}=0.4$ and
$0.5$, respectively.\label{fig:DoS-of-rho in superconductor}}
\end{figure}

As analyzed above, the DoS of $\rho_{S}^{T}$ at temperature $T$
in the conventional superconductor increases monotonically as the
energy gap $\left|\Delta(T)\right|$ decreases. Fig.\ref{fig:DoS-of-rho in superconductor}(a)
shows $\Delta(T)$ in unit of $\Delta(0)$ as a function of $T/T_{c}$.
The energy gap $\Delta(T)$ decreases as $T$ increases and stays
zero when $T>T_{c}$ which means that the system returns to its normal
phase. The temperature dependence of the DoS is shown in Fig.\ref{fig:DoS-of-rho in superconductor}(b).
As can be seen from the figure, the DoS grows as the temperature increases
in the region of $0\leq T\leq T_{c}$, and reaches its maximum value
at the critical temperature $T_{c}$. Then, the increasement of the
temperature above $T_{c}$ does not modify the DoS. In comparison
with the temperature dependent behavior shown by the energy gap, it
follows that the monotonic increasing of the DoS serves as a quantification
for the restoring of the broken symmetry that traditionally depicted
by the decrease of $\Delta(T)$. 

\section{Quantifying SSB in bosons system}

The bosons system can be mapped into the many spin model with the
Hamiltonian

\begin{eqnarray}
H_{B} & = & \xi_{B}\sum_{i=1}^{N}e^{-i\frac{\theta}{2}\sigma_{y}^{i}}\sigma_{z}^{i}e^{i\frac{\theta}{2}\sigma_{y}^{i}},
\end{eqnarray}
where 
\begin{align*}
\xi_{B}=\sqrt{\epsilon^{2}+\lambda^{2}}, & \sin\theta=\lambda/\xi_{B}.
\end{align*}
In the limit of large $N$ and low excitation, the ground state $\left|G\right\rangle _{B}=\prod_{i=1}^{N}e^{-i\frac{\theta}{2}\sigma_{y}^{i}}\left|\downarrow\right\rangle _{i}$
is approximately equivalent to a coherent state, i.e., $\left|G\right\rangle _{B}\simeq\left|\alpha\right\rangle $,
where $\alpha=-\sqrt{N}\theta/2$. It sounds meaningful in physics
that the ground state of BEC is also a coherent state as shown in
Eq.(\ref{eq:BEC coherent}) .

Similar to the case of the BCS theory, we can obtain 
\begin{equation}
S(\mathrm{SO}(2)^{\otimes N},H_{B})=1-\frac{\lambda^{2}}{2\left(\epsilon^{2}+\lambda^{2}\right)}.
\end{equation}

As $\mathrm{Tr}(\left(\rho_{B}^{T=0}\right)^{2})=1$, the DoS of the
ground state reads 
\begin{eqnarray}
S(\mathrm{SO}(2)^{\otimes N},\left|G\right\rangle _{B}) & = & \frac{1}{2}+\frac{1}{2}(1-\frac{\lambda^{2}}{2\xi_{B}^{2}})^{N}.
\end{eqnarray}
The order parameter at $T=0$ can be calculated as 
\begin{align}
\left\langle a\right\rangle _{0} & =\frac{1}{\sqrt{N}}{}_{B}\left\langle G\right|\sum_{i}\sigma_{-}^{i}\left|G\right\rangle _{B}\nonumber \\
 & =-\frac{\sqrt{N}}{2}\sin\theta.
\end{align}
Furthermore, the DoS of the ground state can be re-expressed in term
of the order parameter,

\begin{equation}
S(\mathrm{SO}(2)^{\otimes N},\left|G\right\rangle _{B})=\frac{1}{2}+\frac{1}{2}(1-\frac{2\left\langle a\right\rangle _{0}^{2}}{N})^{N}.\label{eq:DoS GS BEC}
\end{equation}
As $\lim_{x\rightarrow\infty}(1+1/x)^{x}=e$, Eq.(\ref{eq:DoS GS BEC})
is simplified in the limit of large $N$ as

\begin{equation}
\lim_{N/2\left\langle a\right\rangle _{0}^{2}\rightarrow\infty}S(\mathrm{SO}(2)^{\otimes N},\left|G\right\rangle _{B})=\frac{1}{2}+\frac{1}{2}e^{-2\left\langle a\right\rangle _{0}^{2}}.
\end{equation}

This is the second main result of this paper, just as the case for
the conventional superconductor, the maximum of DoS is directly associated
with the vanishing of the order parameter $\langle a\rangle_{0}$.
Thus the DoS of the ground state which is less than unity can also
indicate the SSB in the bosons system. The same results also hold
for the system at finite temperature, since by evaluating Eq.(\ref{eq:DoS rho})
with $\rho_{B}^{T}$ we found

\begin{equation}
S(\mathrm{SO}(2)^{\otimes N},\rho_{B}^{T})=\frac{1}{2}+\frac{\overline{\mathrm{Tr}(R(\omega)\rho R^{\dagger}(\omega)\rho)}_{\mathrm{SO}(2)^{\otimes N}}}{2\mathrm{Tr}\left(\rho^{2}\right)}.
\end{equation}
The DoS of $\rho_{B}^{T}$ of BEC is given as follows 
\begin{eqnarray}
S(\mathrm{SO}(2)^{\otimes N},\rho_{B}^{T}) & = & \frac{1}{2}+\frac{1}{2}\left(1-\frac{\lambda^{2}}{2\xi_{B}^{2}}\frac{\cosh\left(2\beta\xi_{B}\right)-1}{\cosh\left(2\beta\xi_{B}\right)}\right)^{N}.\nonumber \\
\label{eq:BEC T}
\end{eqnarray}
Using the approach similar to the above, we rewritten Eq.(\ref{eq:BEC T})
in terms of the order parameter in a finite temperature $\left\langle a\right\rangle _{T}=\mathrm{Tr}(\rho_{B}^{T}a)=-\lambda\sqrt{N}\tanh(\beta\xi_{B})/(2\xi_{B})$.

\begin{align}
S(SO(2)^{\otimes N},\rho_{B}^{T}) & =\frac{1}{2}+\frac{1}{2}\left[1-\frac{2\left\langle a\right\rangle _{T}^{2}}{N}\frac{1+\cosh\left(2\beta\xi_{B}\right)}{\cosh\left(2\beta\xi_{B}\right)}\right]^{N}.\nonumber \\
\end{align}

Similar to the absolute zero temperature case, the DoS of $\rho_{B}^{T}$
in bosons system depends both on the order parameter $\left\langle a\right\rangle _{T}$
and the temperature $T$. Just like the Penrose-Onsager criterion
\cite{1956 Onsager}, the BEC occurs if and only if $\left\langle a\right\rangle _{T}$
is nonzero. When there exists BEC, a large fraction of particles (to
the order of $N$) occupy the ground state with zero momentum. When
$T>T_{c}$, where $T_{c}$ is the critical temperature of BEC, no
condensation occurs. Thus $\left\langle a\right\rangle _{T>T_{c}}=0$,
and the symmetry of the bosons system is unbroken.

\section{Conclusion}

In this paper, we have exploited a measure of symmetry\textemdash the
degree of symmetry (DoS) to describe the SSB in the conventional superconductor
and BEC. We have established rigorous relations between the DoS and
the order parameters at the absolute zero temperature and finite temperature.
It has been demonstrated that for both the fermions and the bosons
systems, (i) at $T=0$, the order parameter takes its maximum and
the symmetry of the system is maximally broken; (ii) at $0<T<T_{c}$,
the order parameter is still non-vanishing and the extent of the SSB
can be quantified by the DoS; (iii) when $T$ grows beyond $T_{c}$,
the order parameter vanishes and the symmetry of the system is fully
restored.

In fact, the DoS approach that we applied in this paper can be generalized
to other circumstances. We can explore symmetry breaking in other
quantum many-body systems employing the DoS quantifier and expect
to obtain new order parameters when SSB appears. What is worth mentioning
is that the new order parameter must be measurable and reasonable
in physics. 
\begin{acknowledgments}
We thank Yao Yao for helpful discussions. This work was supported
by the National 973 program (Grant No. 2014CB921403), the National
Key Research and Development Program (Grant No. 2016YFA0301201), and
the National Natural Science Foundation of China (Grant Nos.11421063
and 11534002). 
\end{acknowledgments}

\appendix

\section{the DoS of the ground state in the fermions system\label{sec:A}}

As shown in Eq.(\ref{eq:DoS GS fermions 2}), the DoS of the ground
state in the fermions system is

\begin{widetext} 
\begin{align}
S(\mathrm{SO}(2)^{\otimes N},\left|G\right\rangle _{S}) & =\frac{1}{2}+\frac{1}{2}\prod_{k}\left(1-\frac{1}{2}\sin^{2}\theta_{k}\right)\nonumber \\
 & =\frac{1}{2}+\frac{1}{2}\exp\left[\sum_{k}\ln\left(1-\frac{1}{2}\sin^{2}\theta_{k}\right)\right]\nonumber \\
 & =\frac{1}{2}+\frac{1}{2}\exp\left[\int_{-\hbar\omega_{D}}^{\hbar\omega_{D}}g(\epsilon)\ln\left(1-\frac{1}{2}\sin^{2}\theta_{k}\right)d\epsilon\right]\nonumber \\
 & =\frac{1}{2}+\frac{1}{2}\exp\left[\left|\Delta(0)\right|G(\omega_{D},\left|\Delta(0)\right|)\right]\label{eq:Appendix A1}
\end{align}
where 
\[
G(\omega_{D},\left|\Delta(0)\right|)=\int_{-\frac{\hbar\omega_{D}}{\left|\Delta(0)\right|}}^{\frac{\hbar\omega_{D}}{\left|\Delta(0)\right|}}g(t\left|\Delta(0)\right|)\ln\left(1-\frac{1}{2}\frac{1}{t^{2}+1^{2}}\right)dt,
\]
and $\omega_{D}$ is the Debye frequency. As in general case $\hbar\omega_{D}/\left|\Delta(0)\right|\gg1$
and $g(\epsilon)$ changes slowly in the range of $(-\hbar\omega_{D},\hbar\omega_{D})$,
we can take the integral limits to $\pm\infty$ and replace $g(t\left|\Delta(0)\right|)$
with $g(0)$. Therefore, we obtain 
\begin{align}
G(\omega_{D},\left|\Delta(0)\right|) & =\int_{-\frac{\hbar\omega_{D}}{\left|\Delta(0)\right|}}^{\frac{\hbar\omega_{D}}{\left|\Delta(0)\right|}}g(t\left|\Delta(0)\right|)\ln\left(1-\frac{1}{2}\frac{1}{t^{2}+1^{2}}\right)dt\nonumber \\
 & \simeq g(0)\int_{-\infty}^{\infty}\ln\left(1-\frac{1}{2}\frac{1}{t^{2}+1^{2}}\right)dt\nonumber \\
 & =-(2-\sqrt{2})\pi g(0).
\end{align}

In this sense, the DoS of the ground state in the fermions system
is simplified as

\begin{align}
S(\mathrm{SO}(2)^{\otimes N},\left|G\right\rangle _{S}) & =\frac{1}{2}+\frac{1}{2}\exp\left[\left|\Delta(0)\right|G(\omega_{D},\left|\Delta(0)\right|)\right]\nonumber \\
 & \simeq\frac{1}{2}+\frac{1}{2}\exp\left[-(2-\sqrt{2})\pi g(0)\left|\Delta(0)\right|\right]\label{eq:Appendix A2}
\end{align}
\end{widetext}


\begin{thebibliography}{10}
\bibitem{Noether theorem}Noether, E. Invariante variationsprobleme.
\emph{Nachrichten von der Gesellschaft der Wissenschaften zu Gottingen}
\textbf{2}, 235 (1918).

\bibitem{Kibble}T. W. B. Kibble, J. Phys. A: Math. Gen. \textbf{9},
1387 (1976).

\bibitem{Zurek}W. H. Zurek, Nature \textbf{317}, 505 (1985).

\bibitem{Kibble-Zurek}T. W. B. Kibble, Symmetry Breaking and Defects,
in \emph{Patterns of Symmetry Breaking}, Eds. H. Arodz, J. Dziarmaga,
and W. H. Zurek (Springer, Cracow, 2003).

\bibitem{Bunkov}Y. Bunkov, Physica B \textbf{329}, 70 (2003).

\bibitem{mathmaticas}E. H. Lockwood, R. H. Macmillan, \emph{Geometric
Symmetry}, (Cambridge Press, London, 1978).

\bibitem{biology}Y. Saito and H. Hyuga, Rev. Mod. Phys. \textbf{85},
603, (2013).

\bibitem{chemistry}J. P. Lowe, \emph{Quantum Chemistry}, (Academic
Press, Boston, 1993).

\bibitem{SSB}V A Miransky, \emph{Dynamical Symmetry Breaking In Quantum
Field Theories,} (World Scientific, 1994).

\bibitem{2014 Kibble}T. W. B. Kibble, Phil. Trans. R. Soc. A \textbf{373},
33 (2014).

\bibitem{1964 Higgs}P. W. Higgs, Phys. Rev. Lett. \textbf{13}, 508
(1964).

\bibitem{2016 Wang}Z. M. Wang, W. L. Xu, L. Liu, and T. F. Zhu, Nat.
Chem. \textbf{8}, 698 (2016).

\bibitem{1969 Landau}L. D. Landau and E. M. Lifshitz, \emph{Statistical
Physics Volume One}, (Pergamon Press, Oxford, 1969).

\bibitem{1957 BCS}J. Bardeen, L. N. Cooper, J. R. Schrieffer, Phys.
Rev. \textbf{106}, 162 (1957).

\bibitem{1956 Onsager}O. Penrose and L. Onsager, Phys. Rev. \textbf{104},
576 (1956).

\bibitem{ODLRO Yang}C. N. Yang, Rev. Mod. Phys. \textbf{34}, 694
(1962)

\bibitem{2016 Y.N.Fang quantification of nsymmetry}Y. N. Fang, G.
H. Dong, D. L. Zhou, C. P. Sun, Commun. Theor. Phys. \textbf{65}.
423 (2016).

\bibitem{Yao DGH}Yao Yao, G. H. Dong, Xing Xiao, C. P. Sun, arXiv:1605.00789.

\bibitem{Quantum phase transition}S. Sachdev, \emph{Quantum phase
transitions, }(Cambridge University Press, Cambridge, 1999).

\bibitem{1956 Cooper pairs}L. N. Cooper, Phys. Rev. \textbf{104},
1189 (1956).

\bibitem{Kerson Huang}Kerson Huang, \emph{Statistical} \emph{Mechanics},
2nd ed.,Wiley, New York (1987).
\end{thebibliography}
\end{document}